\RequirePackage{fix-cm}
\documentclass{article}                     %
\usepackage{graphicx}

\newcommand{\unused}[1]{}

\newcommand{\del}[1]{}

\usepackage{hyperref}
\hypersetup{colorlinks = true,
            linkcolor = blue,
            urlcolor  = blue,
            citecolor = blue,
            anchorcolor = blue}

\usepackage{natbib}

\begin{document}

\title{{\it This Time with Feeling:} \\
Learning Expressive Musical Performance%
}

\author{Sageev Oore\thanks{Dalhousie University and Vector Institute; work done while author at Google Brain}        \and        Ian Simon\thanks{Google Brain}          \and        Sander Dieleman\thanks{DeepMind}     \and        Douglas Eck\thanks{Google Brain}         \and        Karen Simonyan\thanks{DeepMind}}

\maketitle

{\bf Abstract}:
Music generation has generally been focused on either creating scores or interpreting them. We discuss differences between these two problems and propose that, in fact, it may be valuable to work in the space of direct {\it performance} generation: jointly predicting the notes {\it and also} their expressive timing and dynamics. We consider the significance and qualities of the data set needed for this. Having identified both a problem domain and characteristics of an appropriate data set, we show an LSTM-based recurrent network model that subjectively performs quite well on this task. Critically, we provide generated examples. We also include feedback from professional composers and musicians about some of these examples.


\newpage

\section*{Preamble/Request}

 Recognizing that ``talking about music is like dancing about architecture''\footnote{This quote has been attributed to a range of individuals from Laurie Anderson to Miles Davis, and numerous others.}, we kindly ask the reader to listen to the linked audio in order to effectively understand the motivation, data, results, and conclusions of this paper. As this research is ultimately about producing music, we believe the actual results are most effectively perceived---indeed, only perceived---in the audio domain.
 This will provide necessary context for the verbal descriptions in the rest of the paper.

\section{Introduction}

In this work, we discuss training a machine-learning system to generate music. The first two key words in the title are {\it time} and {\it feeling}: not coincidentally, our central thesis is that, given the current state of the art in music generation systems, it is effective to generate the expressive timing and dynamics information concurrently with the music. Here we do this by directly generating improvised performances rather than creating or interpreting scores. %
We begin with an exposition of some relevant musical concepts.

\subsection{Scores, Performances and Musical Abstraction}
\label{s:abstraction}

Music exists in the audio domain, and is experienced through individuals' perceptual systems. Any ``music'' that is not in the audio domain (e.g. a text or binary file of any sort) is of course a representation of music: if it is not physically vibrating, it is not (yet) sound, and if it is not sound, it is certainly not music. 
The obvious implication is that for any representation, there are additional steps to transform that representation---whatever it might be---into sound. Those steps might be as local as the conversion from digital to analog waves, or as global as the human performance of written score, for example. In generating\del{?} music\footnote{In this text, we use the term ``generation'' to refer to computational generation, as opposed to human creation or performance.}, therefore, one must be aware of which of those steps is addressed directly by their generative system, which ones must be addressed in other ways, and, importantly, the impact of all of those choices on the listener's perception of the music, where it is ultimately experienced.

A defining characteristic of a representation, then, is what is omitted: what still needs to be added or done to it in order to create music from it, and the relation of that abstraction to our perceptual experience. With that consideration in mind, we now discuss some common symbolic representations.

\subsubsection{Scores \del{What is a score?}}
\label{s:score}

\begin{figure}
\includegraphics[scale=0.45]{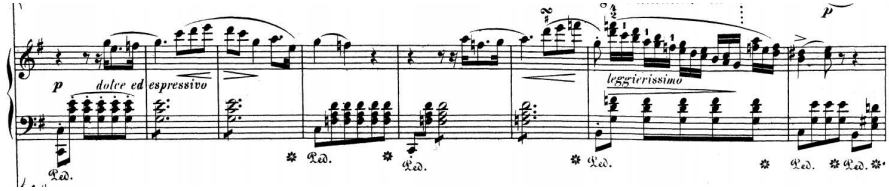}
\caption{Excerpt from the score of Chopin's Piano Concerto No. 1.}
\label{f:chopin1}
\end{figure}

Figure~\ref{f:chopin1} is an example of a musical score~\cite{chopin-1830}. It shows which notes to play and when to play them relative to each other. The timing in a score is aligned to an implicit and relative {\it metrical grid}. For example, quarter notes are the same duration as quarter note rests, twice the duration of eighth notes, and so on. Some scores additionally specify an absolute tempo, e.g. in quarter notes per minute. 

And yet, by the time the music is heard as audio, most of this timing information will have been intentionally {\it not followed exactly}! For example, in classical music from the 1800's onwards, {\it rubato} developed: an expressive malleability of timing that overrides metrical accuracy (i.e. can deviate very far from the grid), and this device is both frequent and essential for making perceptual sense of certain pieces. Another example of a rhythmic construct that is not written in scores is {\it swing}, a defining quality of many African American \del{check footnote here} music traditions\footnote{While explaining swing is outside the current scope, we do note that it is occasionally incorrectly described in terms of triplets. %
}.

But tempo is not the only way in which the score is not followed exactly. {\it Dynamics} refers to how the music gets louder and quieter. While scores do give information about dynamics, in this respect, too, their effectiveness relies heavily on conventions that are not written into the score. For example, where the above score says ``{\it p}'' it means to play quietly, but that does tell us how quietly, nor will all the notes be equally quiet. When there is a {\it crescendo} marking indicating to get louder, in some cases the performer will at first get momentarily quieter, creating space from which to build. Furthermore, when playing polyphonic piano music, notes played at the same time will usually be played at different dynamic levels and articulated differently from one another in order to bring out some voices over others.

{\it Phrasing} includes a joint effect of both expressive timing and dynamics. For example, there is a natural correlation between the melody rising, getting louder, and speeding up. These are not rules, however; skilled performers may deliberately choose to counteract such patterns to great effect.

We can think of a score as a highly abstract representation of music. The effective use of scores, i.e. the assumption by a composer that a score will subsequently be well-rendered as music, relies on the existence of conventions, traditions, and individual creativity. For example, Chopin wrote scores where the pianist's use of {\it rubato} is expected, indeed the score requires it in order to make sense. Similarly, the melodies in jazz lead sheets were written with the understanding that they will be {\it swung} and probably embellished in various ways. There are numerous other instrument-specific aspects that scores do not explicitly represent, from the vibrato imbued by a string player to the tone of a horn player. Sometimes, the score just won't really make perceptual sense without these.

In short, the mapping from score to music is full of subtlety and complexity, all of which turns out to be very important in the perceptual impact that the music will have. To get a sense of the impact of these concepts, we recommend that the reader listen:
\begin{itemize}
\item first to a {\it direct rendering} of the above score\del{ask editor about links} here: \url{https://clyp.it/jhdkghso}, played according to the written grid and quantized to $16^{th}$ notes. Then, 
\item listen to an {\it expressive performance}~\cite{oore-chopinPerf} of it here: \url{https://clyp.it/x24hp1pq}.
\end{itemize}

\subsubsection{MIDI \del{What is MIDI?}}
\label{s:piano-roll}

\begin{figure}
\includegraphics[scale=0.35]{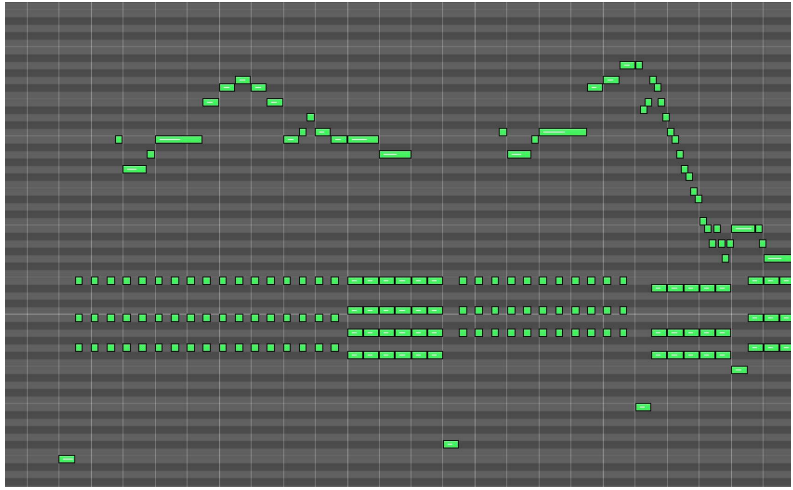}
\caption{Piano roll based on the score in Figure~\ref{f:chopin1}. The horizontal axis represents time; the vertical axis represents pitch; each rectangle is a note; and the length of the rectangle corresponds to the duration of the note.}
\label{f:chopin-roll}
\end{figure}

MIDI is a communication protocol for digital musical instruments: a symbolic representation, transmitted serially, that indicates {\sc Note\_On} and {\sc Note\_Off} events and allows for a high temporal sampling rate. The loudness of each note is encoded in a discrete quantity referred to as {\it velocity} (the name is derived from how fast a piano key is pressed). While MIDI encodes note timing and duration, it does not encode qualities such as timbre; instead, MIDI events are used to trigger playback of audio samples.

MIDI can be visualized as a piano roll---a digital version of the old player piano rolls.
Figure~\ref{f:chopin-roll} is an example of a MIDI piano roll corresponding to the score shown in Figure~\ref{f:chopin1}. Each row corresponds to one of the 128 possible MIDI pitches. Each column corresponds to a uniform time step. If note $i$ is ON at time $t$ and had been pressed with velocity $v$, then element $(t, i) = v$. So, at 125~Hz, six seconds of MIDI data would be represented on a grid of size $128 \times (6 \times 125)$. Actual MIDI sampling can be faster than this, so even at 125~Hz we are still subsampling from the finest available temporal grid. 

We refer to a score that has been rendered directly into a MIDI file as a {\bf{MIDI Score}}. That is, it is rendered with no dynamics and exactly according to the written metrical grid. As given earlier, \url{https://clyp.it/jhdkghso} is an example of this.

If, instead, a score has been performed, by a musician for example, and that performance has been encoded into a MIDI stream, we refer to that as a {\bf{MIDI Performance}}. \url{https://clyp.it/x24hp1pq} is an example (also given previously) of a MIDI performance.

\section{Factoring the Music Generation Process: Related Work}
\label{s:domain}
\label{s:related}

\begin{figure}
\includegraphics[trim={0cm 1cm 0cm 5cm},clip,scale=0.45]{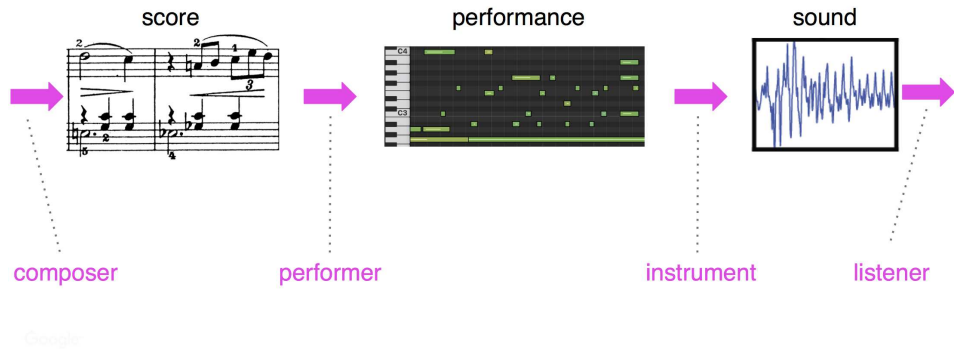}
\caption{Factoring music generation. We can see music as starting with the composition of a score; that score gets turned into a performance (shown as a MIDI piano roll); that MIDI roll, in turn, gets rendered into sound using a synthesizer, and finally the resulting audio gets perceived as music by a human listener. }
\label{f:factoringMusicGen1}
\end{figure}

Figure~\ref{f:factoringMusicGen1} shows one way of factoring the music generation process. The first stage shown in this figure is composition, which yields a score. The score is then performed. The performance is rendered as sound, and finally that sound is perceived. In the analog world, of course, performance and rendering the sound are the same on a physical instrument, but in the digital world, those steps are often separate. While other views of the process are possible, this one provides us a helpful context for considering much of the existing relevant work. Noting that sound generation and perception (the last two steps in Figure~\ref{f:factoringMusicGen1}) are outside our scope, in the rest of this section we focus primarily on composition and performance.

Perhaps it is precisely because music is so often perceived as a profoundly human endeavour that there has also been, in parallel, an ongoing fascination with automating its creation. This fascination long predates notions such as the Turing test (ostensibly for discriminating automation of the most human behaviour), and has spawned a range of efforts: from attempts at the formalization of unambiguously strict rules of composition to incorporation of complete random chance into scores and performances. The use of rules exemplifies the algorithmic (and largely deterministic) approach to music generation, one that is interesting and outside the scope of the current work; for background on this we refer the reader, for example, to the text by Nierhaus~\cite{nierhaus-2009}. Our present work, on the other hand, lies in a part of the spectrum that incorporates probability and sampling.

{\bf {\it Aleatory}} refers to music or art that involve elements of randomness, derived from the Latin {\it alea (alee)}, meaning ``die (dice)''. Dice were used in the 1700's to create music in a game referred to as {\it Musikalisches W\"{u}rfelspiel}~\cite{nierhaus-2009,hedges-78,boehmer-67}: the rolled numbers were used to select from pre-composed fragments of music. Some of these compositions were attributed to Mozart and Haydn, though this has not been authenticated. 

Two centuries later, as the foundations of AI were being set, the notion of automatically understanding (and therefore generating) music was among the earliest applications to capture the imagination of researchers,  with papers on computational approaches to perception, interpretation and generation of music by Simon, Longuet-Higgins and others~\cite{linblom-sundberg-70,longuet-higgins-76,longuet-higgins-78,longuet-higgins-steedman-71,simon-sumner-68}.  
Since then, many interesting efforts were made~\cite{griffith-todd,todd-loy,concert-94,eck-schmidhuber-2002,pachet-2003,hild-91},
and it is clear that in recent years both interest and progress in score generation has continued to advance, e.g. Lattner et al~\cite{lattner-2017}, Boulanger-Lewandowski et al~\cite{boulanger-lewandowski-et-al-2012}, Bretan et al~\cite{bretan-2017a}, Herremans et al~\cite{herremans-chew-2017}, Roberts et al~\cite{roberts-2016}, Sturm~\cite{sturm-2016}, to name only a few. Briot et al~\cite{briot-2017} provide a survey of generative music models that involve machine learning. Herremans et al~\cite{herremans-chuan-chew-2017} provide a comprehensive survey and satisfying taxonomy of music generation systems. McDonald~\cite{mcdonald-2017} gives an overview highlighting some key examples of such work.

Corresponding to the second step in Figure~\ref{f:factoringMusicGen1} is a body of work often referred to as EMP (Expressive Musical Performance) systems. For example, the work by Chacon and Grachten~\cite{chacon-grachten-2016}, inspired by the Linear Basis Models proposed by Grachten and Widmer~\cite{grachten-widmer-2012}, involves defining a set of hand-engineered features, some of which depend on having a score with dynamic expression marks, others on heuristics for musical analysis (e.g. a basis function indicating whether the note falls on the first beat of a measure of 4/4). \del{\tiny{While hand-developed features were certainly very important in acquiring a more thorough understanding of the problem domain, this differs from our work in that we consider only the MIDI signal itself: velocities and the near-exact timing of $NOTE\_ON$ and $NOTE\_OFF$ events. Any features beyond that are computed and represented implicitly by the system. }}
Widmer and Goebl~\cite{widmer-goebl-2004} and Kirke and Miranda~\cite{kirke-miranda-2013} both present extensive and detailed surveys of work done in the field of computational EMPs. \del{could move these survey sentences towards the beginning of this paragraph} In the latter survey, the authors also provide a tabular comparison of 29 systems that they have reviewed. Out of those systems, two use neural networks (one of which also uses performance rules) and a few more use PCA, linear regression, KCCA, etc. Some of the other systems that involve some learning, do so by learning rules in some way. For example, the KTH model~\cite{friberg-et-al-2006} consists of a top-down approach for predicting performance characteristics from rules based on local musical context. Bresin~\cite{bresin-1998} presents two variations of a neural network-based system for learning how to add dynamics and timing to MIDI piano performance.%

Grachten and Krebs~\cite{grachten-krebs-2014} use a variety of unsupervised learning techniques to learn features with which they then predict expressive dynamics. Building on that work, van Herwaarden et al~\cite{vanherwaarden-et-al-2014} use an interesting combination of an RBM-based architecture, a note-centered input representation, and multiple datasets to---again---predict expressive dynamics. In both of these cases, the dynamics predictions appear to depend on the micro-timing rather than being predicted jointly as in the present work.

Teramura et al~\cite{teramura-et-al-2008}, observe that many previous performance rendering systems ``often consist of many heuristic rules and tend to be complex. It makes [it] difficult to generate and select the useful rules, or perform the optimization of parameters in the rules.''
They thus present a method that uses Gaussian Processes to achieve this, where some parameters can be learned. In their ostensibly simpler system, ``for each single note, three outputs and corresponding thirteen input features are defined, and three functions each of which returns one of three outputs and receive the thirteen input features, are independently learned''. However, some of these features, too, depend on certain information, e.g. they compute the differences between successive pitches, and this only works in compositions where the voice leading is absolutely clear; in the majority of classical piano repertoire, this is not the case. In Laminae~\cite{okumura-sako-kitamura-2014}, Okumura et al systematize a set of context-dependent models, building a decision tree which allows rendering a performance by combining contextual information.

Moulieras and Pachet~\cite{moulieras-pachet-2016} use a maximum entropy model to generate expressive music, but their focus is again monophonic plus simple harmonic information. They also explicitly assume that ``musical expression consists in local texture, rather than long-range correlations''. While this is fairly reasonable at this point, and indeed it is hard to say how much long-range correlation is captured by our model, we wished to choose a model which, at least in principle, allowed the possibility of modeling long-range correlation: ultimately, we believe that these correlations are of fundamental importance. Malik and Ek~\cite{malik-ek-2017} use a neural network to learn to predict the dynamic levels of individual notes while assuming quantized and steady timing.

\section{Choosing Assumptions and a Problem Domain}
\label{s:domain}

\subsection{Assumptions}

In the case of both score production and interpretation, any computational model naturally makes assumptions. Let us review potential implications of some of these when generating music, and identify some of the choices we make in our own model in these respects.

\begin{itemize}
    \item {\bf Metric Abstraction} %
    Many systems abstract rhythm in relation to an underlying grid, with metric-based units such as eighth notes and triplets. Often this is further restricted to step sizes at powers of two.  Such abstraction is oblivious to many essential musical devices, including e.g. rubato and swing as described in Section~\ref{s:score}. Some EMP systems allow for variations in the global tempo, %
    but this would not be able to represent common performance techniques such as playing of the melody slightly staggered from accompaniment (i.e. creating an asynchrony beyond what is written in the score). \del{It is worth noting that when a composer writes a piece where the melody might be performed deliberately asynchronously with the accompaniment, the composer is generally well-aware that this is within the musical vocabulary of the performer.} 
    
    {\it We choose a temporal representation based on absolute time intervals between events, rounded to 8ms.}\\
    
    \item {\bf No Dynamics} %
    Nearly every compositional system represents notes as ON or OFF. This binary representation ignores dynamics,
    which constitute an essential aspect of how music is perceived.
    The EMP systems do tend to focus on dynamics. While many systems do not have audio readily available, we point out that listening to, e.g. the work of Malik and Ek~\cite{malik-ek-2017} where a binned velocity value is predicted for each note, the abstracted and static tempo is still quite noticeable. When dynamic level is treated in some EMPs as a global parameter applied equally to simultaneous notes, this defeats the ability of dynamics to differentiate between voices, or to compensate for a dense accompaniment (that is best played quietly) underneath a sparse melody. 
    \\
    {\it We allow each note to have its own dynamic level.}\\
    
    \item {\bf Monophony} %
    Some systems only generate monophonic sequences.  Admittedly, one must start somewhere: the need to limit to monophonic output is in this sense entirely understandable. This can work very well for instruments such as voice and violin, where the performer also has sophisticated control beyond quantized pitch and the velocity of the note attack. The perceived quality of monophonic sequences may be inextricably tied to these other dimensions that are difficult to capture and usually absent from MIDI sequences.
    
    In our experience, the leap from monophonic to polyphonic generation is a significant one. A survey of the literature shows that most systems that admit polyphony still make assumptions about its nature---either that it is separable into chords, or that it is separable into voices, or that any microvariation in tempo applies to all voices at once (as opposed to allowing one voice to come in ahead of the beat), and so forth. Each of these assumptions is correct only sometimes. We  settled on a representation that turned out to be simpler and more agnostic than this, in that it does not make any of these assumptions: \\
    {\it We specify note events one at a time, but allow the system to predict an arbitrary number of simultaneous notes, should it be so inclined.}\\

\end{itemize}

Generally speaking, in contrast to many of the method discussed in Section~\ref{s:related}, our approach makes no assumptions about the features other than the information that is known to exist in MIDI files: velocity, timing and duration of each note. We do not require computing or knowing the time signature, we do not require knowing the voice leading, we do not require inferring the chord, and so on. While additional information could be both useful and interesting, given the current state of the art and available data, we are focused on showing how much can be done without defining any rules or heuristics at all; we simply try to model the distribution of the existing data. Listening to some of the examples, one hears that our system generates a variety of natural time feels, including 3/4, 4/4 and odd time signatures, and they never feel rhythmically heavy-handed.

\subsection{Problem Domain: Simultaneously Composing and Performing}

\begin{figure}
\includegraphics[trim={0cm 4.25cm 0cm 1.5cm},clip,scale=1.05]{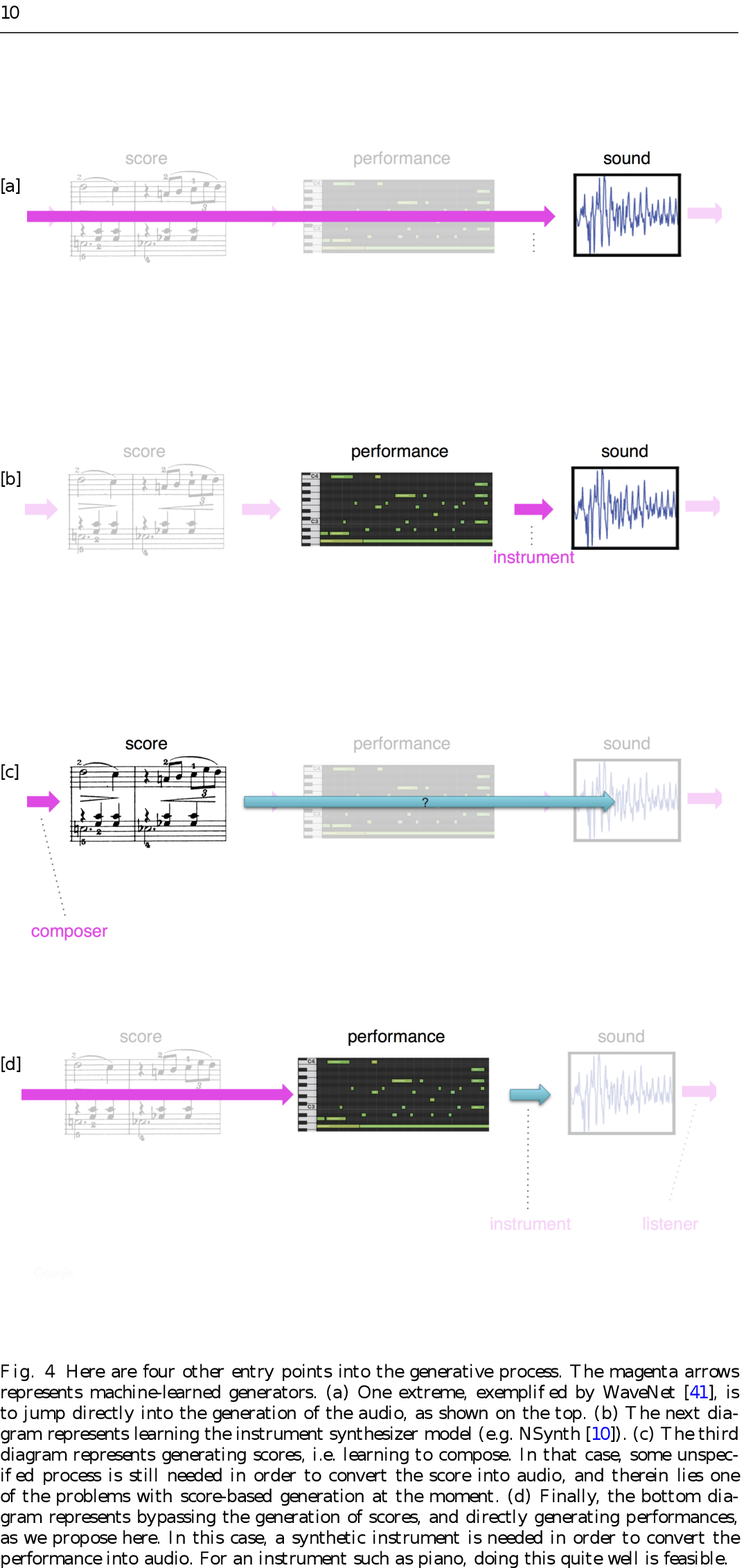}
\vskip-0.5in
\caption{Here are four other entry points into the generative process. The magenta arrows represents machine-learned generators. (a) One extreme, exemplified by WaveNet~\cite{van-den-oord-et-al-2016}, is to jump directly into the generation of the audio, as shown on the top. (b) The next diagram represents learning the instrument synthesizer model (e.g. NSynth~\cite{engel-et-al-2017}). (c) The third diagram represents generating scores, i.e. learning to compose. In that case, some unspecified process is still needed in order to convert the score into audio, and therein lies one of the problems with score-based generation at the moment. (d) Finally, the bottom diagram represents bypassing the generation of scores, and directly generating performances, as we propose here. In this case, a synthetic instrument is needed in order to convert the performance into audio. For an instrument such as piano, doing this quite well is feasible.}
\label{f:factoringMusicGen}
\end{figure}

In Figure~\ref{f:factoringMusicGen}, we show a few different possible entry points to the music generation process. For example, at one extreme, we can subsume all steps into a single mechanism so as to predict audio directly, as is done by WaveNet, with impressive results~\cite{van-den-oord-et-al-2016}.
Another approach is to focus only on the instrument synthesis aspect~\cite{engel-et-al-2017}, which is an interesting problem outside the scope of our present work. As described in Section~\ref{s:related}, the compositional systems generate scores that require performances, while the EMP systems require scores in order to generate performances.

Here, we demonstrate that {\it jointly predicting composition and performance with expressive timing and dynamics}, as illustrated in Figure~\ref{f:factoringMusicGen}(d), is another effective domain for music generation given the current state of the art. Furthermore, it creates output that can be listened to without requiring additional steps beyond audio synthesis as provided by a piano sample library.

While the primary evidence for this will be found simply by listening to the results, we mention two related discussion points about the state of the art:

\begin{itemize}
\item {\it Music with very long-term, fully coherent structure is still elusive.}  
In ``real'' compositions, long-term structure spans the order of many minutes and is coherent on many levels.  There is no current system that is able to learn such structure effectively. That is, if $\Delta t = 8ms$, then even for just 2 minutes, $P( e_{i + 15000} | e_{i} )$ should be different from $P( e_{i + 15000})$. There is no current system that effectively achieves anywhere near this for symbolic MIDI representation. 

\item {\it Metrics for evaluating generated music are very limited.}  Theis and others~\cite{theis-et-al-2016,van-den-oord-dambre-2015} have given clear arguments about the limitations of metrics for evaluating the quality of generative models in the case of visual models, and their explanations extend naturally to the case of musical and audio models. In particular, they point out that ultimately, ``models need to be evaluated directly with respect to the application(s) they were intended for''. In the case of the generative music models that we are considering, this involves humans listening.

\end{itemize}

Taken together, what this means is that systems that generate musical scores face a significant evaluation dilemma. Since by definition any listening-based evaluation must operate in the audio space, either a) the scores must be rendered directly and will lack expression entirely, or b) a human or other system must perform the scores, in which case the quality of the generated score is hard to disentangle from the quality of the performance.\footnote{For example, listening to the direct score and performance clips given above, it should be clear that other than perhaps very experienced musicians, it would be extremely difficult for a listener to hear the audio of the MIDI Score and intuitively understand that that same passage could sound as it does in the MIDI Performance.} Furthermore, the lack of long-term structure compounds the difficulty of evaluation, because one of the primary qualities of a good score is precisely in its long-term structure. This implicitly bounds the potential significance of evaluating a short and context-free compositional fragment.

With these considerations in mind, we generate directly in the domain of musical performance.  A side benefit of this is that informal evaluation becomes more potentially meaningful: musicians and non-musicians alike can listen to clips of generated performances while (1) not being put off by the lack of expressiveness and (2) not needing to disentangle the different elements that contributed to what they hear, since both the notes and how they are all played were all generated by the system.\footnote{We emphasize that these observations do not apply to the development of tools for composers, where score fragment generation might be appropriate. Also, we reiterate that this discussion is made in relation to the current state of the art.} We also note that our approach is consistent with many of the points and arguments recently made by Widmer~\cite{widmer-16}.

\section{Data}
\label{s:data}
If we wish to predict expressive performance, we need to have the appropriate data. We use the International Piano-e-Competition dataset~\cite{piano-competition}, which contains MIDI captures of roughly 1400 performances by skilled pianists. The pianists were playing a Disklavier, which is a real piano that also has internal sensors that record MIDI events corresponding to the performer's actions.  The critical importance of good data is well-known for machine learning in general, but here we note some particular aspects of this data set that made it well-suited for our task.

\subsection{Homogeneous}

The data set was homogeneous in a set of important ways. It might be easy to underestimate the importance of any of the following criteria, and so we list them all explicitly here with some discussion:

\paragraph{First, it was all classical music.} This helps the coherence of the output.

\paragraph{Second, it was all solo instrumental music.} If one includes data that is for two or more instruments, then it no longer makes sense to train a generative model that is expected to generate for a solo instrument; there will be many (if not most) passages where what one instrument is doing is entirely dependent on what the other instrument is doing. The text analogy would be hoping for a system to learn to write novels by training it on only one character's dialogue from movies and plays. There will occasionally be self-sufficient monologues, but generally speaking,  well-written dialogue has already been distilled by the playwright, and makes more sense when voices are not removed from it.

\paragraph{Third, that solo instrument was consistently piano.} Classical composers generally write in a way that is very specific to whichever instrument they are writing for. Each instrument has its own natural characteristics, and classical music scores (i.e. that which is captured in the MIDI representation) are very closely related to the timbre of that instrument (i.e. how those notes will be ``rendered''). One exception to this is that Bach's music tends to sound quite good on any instrument, e.g. it is OK to train a piano system on Bach vocal chorales.

\paragraph{Fourth, the piano performances were all done by humans.} The system did not have to contend with learning from a dataset where some of the examples were synthesized, some were ``hand-synthesized'' to appear like human performances, etc. Each of those classes has its own patterns of micro-timing and dynamics, and each may be well-suited for a variety of music-related tasks, but for training a system on performances, it is very helpful that all the performances are indeed$\ldots$ performances.

\paragraph{Finally, all of those humans were experts.} If we wish the system to learn about human performance, that human performance must match the listener's concept of what ``human performance'' sounds like, which is usually performances by experts. The casual evaluator might find themselves slightly underwhelmed were they to listen to a system that has learned to play like a beginning pianist, even if the system has done so with remarkable fidelity to the dynamic and velocity patterns that occur in that situation.

\subsection{Realizable}

The fact that the solo instrument was piano had additional advantages.
Synthesizing audio from MIDI can be a challenging problem for some instruments. For example, having velocities and note durations and timing of violin music would not immediately lead to good-sounding violin audio at all. The problems are even more evident if one considers synthesizing vocals from MIDI. That the piano is a percussive instrument buys us an important benefit: synthesizing piano music from MIDI can sound quite good. Thus, when we generate data we can properly realize it in audio space and therefore have a good point of comparison. 
Conversely, capturing the MIDI data of piano playing provides us with a sufficiently rich set of parameters that we can later learn enough in order to be able to render audio. Note that with violin or voice, for example, we would need to capture many more parameters than those typically available in the MIDI protocol in order to get a sufficiently meaningful set of parameters for expressive performance.

\section{RNN Model}

We modeled the performance data with an LSTM-based Recurrent Neural Network. The model consisted of three layers of 512 cells each, although the network did not seem particularly sensitive to this hyperparameter. We used a temporally non-uniform representation of the data, as described next.

\subsubsection{Representation: Time-shift}

A MIDI excerpt is represented as a sequence of events from the following vocabulary of 413 different events:
\begin{itemize}
    \item {\bf 128 NOTE-ON} events: one for each of the 128 MIDI pitches. Each one starts a new note.
    \item {\bf 128 NOTE-OFF} events: one for each of the 128 MIDI pitches.  Each one releases a note.
    \item {\bf 125 TIME-SHIFT} events: each one moves the time step forward by increments of 8 ms up to 1 second.
    \item {\bf 32 VELOCITY} events: each one changes the velocity applied to all subsequent notes (until the next velocity event).
\end{itemize}

The neural network operates on a one-hot encoding over this event vocabulary. Thus, at each step, the  input to the RNN is a single one-hot 413-dimensional vector. For the piano-e-competition dataset, a 15-second clip typically contains ~600 such one-hot vectors, although this varies considerably (and roughly linearly with the number of notes in the clip).

While the minimal time step is a fixed absolute size ($8~ms$), the model can skip forward in time to the next note event.  Thus, any time steps that contain rests or simply hold existing notes can be skipped with a single event.  The largest possible single time shift in our case is 1 second but time shifts can be applied consecutively to allow effectively longer shifts.  The combination of fine quantization and time-shift events helps maintain expressiveness in note timings while greatly reducing sequence length compared to an uncompressed representation.

\begin{figure}
\includegraphics[trim={0cm 0cm 0cm 0cm},clip,scale=0.5]{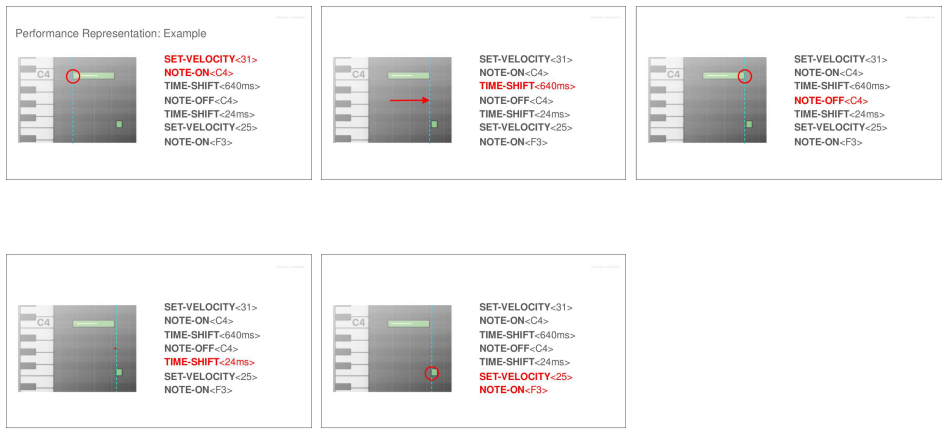}
\caption{Example of Representation used for PerformanceRNN. The progression illustrates how a MIDI sequence (e.g. shown as a MIDI roll consisting of a long note followed by a shorter note) is converted into a sequence of commands (on the right hand side) in our event vocabulary. Note that an arbitrary number of events can in principle occur between two time shifts.}
\label{f:performanceRepresentation}
\end{figure}

This fine quantization is able to maintain expressiveness in note timings while not being as sparse as a grid-based representation. This sequence representation uses more events in sections with higher note density, which matches our intuition.

\begin{figure}
\includegraphics[trim={2cm 0cm 0cm 0cm},clip,scale=0.5]{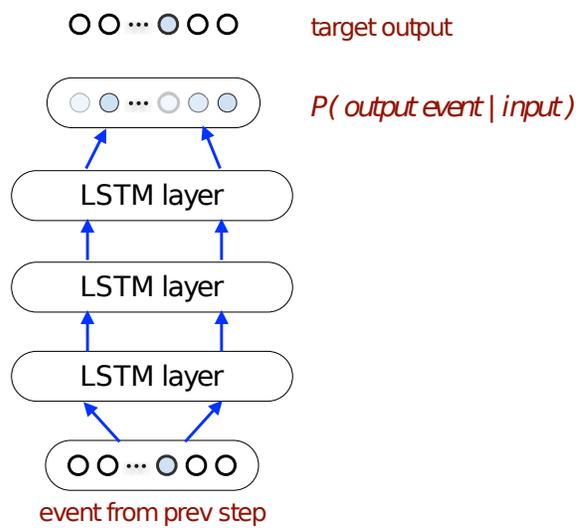}
\caption{The basic RNN architecture consists of three hidden layers of LSTMs, each layer with 512 cells. The input is a 413-dimensional one-hot vector, as is the target, and the model outputs a categorical distribution over the same dimensionality as well. For generation, the output is sampled stochastically with beam search, while teacher forcing is used for training.}
\label{f:perfRNN}
\end{figure}

Figure~\ref{f:performanceRepresentation} shows an example of a small excerpt of MIDI performance data converted to our representation. Figure~\ref{f:perfRNN} shows a diagram of the basic RNN architecture.

\subsection{Training and Data Augmentation}
\label{ss:training-and-data-augmentation}

We train the models by first separating the data into 30-second clips, from which we then select shorter segments. We train using stochastic gradient descent with a mini-batch size of 64 and a learning rate of 0.001 and teacher forcing.

\subsubsection{Augmentation}

We augment the data in two different ways, for different runs:\\

\noindent Less augmentation:
\begin{itemize}
\item Each example is transposed up and down all intervals up to a major third, resulting in 8 new examples plus the original. 
\item Each example is stretched in time uniformly by $\pm 2.5\%$ and $\pm 5\%$, resulting in 4 new examples plus the original.
\end{itemize}

\noindent More augmentation:
\begin{itemize}
\item Each example is transposed up and down all intervals up to 5 or 6 semitones to span a full octave, resulting in 11 new examples plus the original. 
\item Each example is stretched in time uniformly by up to $\pm 10\%$.
\end{itemize}

\subsubsection{Quantization}
\label{ss:quantization}

In Section \ref{s:domain}\del{CHECK THIS SECTION#} we describe several forms of quantization that can be harmful to perceived musical quality.  Our models also operate on quantized data; however, unlike much prior work we aim for quantization levels that are below noticeable perceptual thresholds.   
\begin{description}%
    \item[Timing] \hfill \\
    Friberg and Sundberg~\cite{friberg-sundberg-1992} found that the just noticeable difference (JND) when temporally displacing a single tone in a sequence was generally no finer than roughly $10ms$. Other studies have found that the JND for change in tempo is no finer than roughly 5\%. We note that for a tempo of $120bpm$, each beat lasts for $500ms$, and therefore this corresponds to a change of roughly $25ms$. Given that at that tempo beats will frequently still be subdivided into 2 or triplets, that would correspond to a change of roughly 8~ms per subdivided unit. We therefore assume that using a sampling rate of $125Hz$ (i.e. $1000/8$) should generally be below the typical perceptual threshold.
    \item[Dynamics] \hfill \\
    Working with piano music, we have found that 32 different ``steps'' of velocity are sufficient. Note that there are about 8 levels of common dynamic marking in classical music (from {\it ppp} to {\it fff}), so it may well be the case that we could do with fewer than 32 bins, but our objective was not to find the lower bound here. 

\end{description}

\subsubsection{Predicting Pedal}
\label{predicting-pedal}
 
In the RNN model, we experimented with predicting sustain pedal.  
We applied {\sc Pedal\_On} by directly extending the lengths of the notes: for any notes on during or after a {\sc Pedal\_On} signal, we delay their corresponding {\sc Note\_Off} events until the next {\sc Pedal\_Off} signal. This made it a lot easier for the system to accurately predict a whole set of {\sc Note\_Off} events all at once, as well as to predict the corresponding delay preceding this. Doing so may have also freed up resources to focus on better prediction of other events as well. Finally, as one might expect, including pedal made a significant subjective improvement in the quality of the resulting output.

\section{Results}

We begin with the most important indicator of performance: generated audio examples.

\subsection{Examples}

In these examples, our systems generated all MIDI events: timing and duration of notes as well as note velocities. We then used freely-availably piano samples to synthesize audio from the resulting MIDI file.

A small set of examples are available at \url{https://clyp.it/user/3mdslat4}. We strongly encourage the reader to listen. These examples are representative of the general output of the model. We comment on a few samples in particular, to give a sense of the kind of musical structure that we observe:
\begin{itemize}
\item \href{https://clyp.it/p0qh4hvt}{\color{blue} RNN Sample 4}: This starts off with a slower segment that goes through a very natural harmonic progression in G minor, pauses on the dominant chord, and then breaks into a faster section that starts with a G major chord, then passes through major chords related to G minor (Bb, etc). Harmonically, this shows structural coherence even while the tempo and feel shift. At around 12s, the ``left hand'' uses dynamics to bring out an inner voice in a very natural and appropriate way.
\item \href{https://clyp.it/nytzvxmx}{\color{blue} RNN Sample 7}: This excerpt begins very reminiscent of a Schubert Impromptu, although it is sufficiently different that it has clearly not memorized it. There is a small rubato at the very beginning of the phrase, especially on the first note, which is musically appropriate. The swells in the phrasing make musical sense, as do the slight pauses right before some of the isolated notes in the left hand (e.g. the E at 0:10s, the F$\sharp$ at around 12.5 seconds).
\item \href{https://clyp.it/4pmf2b4p}{\color{blue} RNN Sample 2}: This excerpt begins in a classical style (e.g. Haydn or Mozart). Interestingly, the same way that one note (an F) is repeated in the right hand in the first few seconds, after a pause, the next phrase begins and then at around 8 seconds, the left hand mirrors that articulation pattern with a set of descending repeated notes (A$\flat$, G, F).
\end{itemize}

\subsection{Log-likelihood}

We begin by noting that objective evaluation of these kinds of generative is fundamentally very difficult, and measures such as log-likelihood can be quite misleading~\cite{theis-et-al-2016}. %
Nevertheless, we provide comparisons here over several different hyperparameter configurations for the RNN.

\begin{table}
\centering
\begin{tabular}{|c|c|l|}
\hline
model & log-loss & description \\
\hline
RNN & .765 & baseline RNN trained on 15-second clips \\
RNN-NV & .619 & baseline without velocity \\
RNN-SUS & .663 & baseline with pedaled notes extended \\
RNN-AUG+ & .755 & baseline with more data augmentation \\
RNN-AUG- & .784 & baseline with less data augmentation \\
RNN-30s & .750 & baseline trained on 30-second clips \\
RNN-SUS-30s & .664 & baseline + pedal + 30-second clips \\
\hline
\end{tabular}
\caption{Log-loss of RNN model variants trained on the Piano-e-competition performance dataset and evaluated on a held-out subset.}
\label{rnn-results}
\end{table}

Table~\ref{rnn-results} contains the per-time-step log-loss of several RNN model variants.  The baseline model is trained on 15-second performance clips, ignoring sustain pedal and with the two forms of data augmentation described in Section~\ref{ss:training-and-data-augmentation}.

Note that while RNN-NV has the best log-loss, this variant is inherently easier as the model does not need to predict velocities.  In the RNN-SUS variant, sustain pedal is used to extend note durations until the pedal is lifted; this aids prediction as discussed in Section~\ref{predicting-pedal}.

\subsection{Informal Feedback From Professional Composers and Musicians}

We gave a small set of clips to professional musicians and composers for informal comments. We were not trying to do a Turing test, so we mentioned that the clips were generated by an automated system, and simply asked for any initial reactions/comments. Here is a small, representative subset of the comments we received (musical background in bold, some particularly interesting excerpts are italicized for later discussion):
\\
\\
\noindent {\bf TV/Film composer}: \\
\begin{quote}
Fantastic!!!! How many hours of learning [$\ldots$] here?
\end{quote}
\begin{quote}
This [$\ldots$] absolutely blows the stuff I've heard online out of the solar system.
The melodic sense is still foggy, in my view, but it's staggering that it makes nice pauses with some arcing chord progressions quite nicely. 
I think that it's not far from actually coming up with a worthwhile melody. [$\dots$]
How does it know what ``inspirational emotion'' to draw from? or is it mostly doing things ``in the likeness of''?
\end{quote}
\begin{quote}
Fascinating!!
\end{quote}
\vspace{0.25in} %
\noindent {\bf Composer \& Professional Musician}\\
\begin{quote}
{\it{In terms of performance I'm quite impressed with the results.}} It sounds more expressive than any playback feature I’ve worked with when using composition software.\\
\end{quote}
\begin{quote}
{\it{In terms of composition, I think there is more room for improvement.}} The main issue is lack of consistency in rhythmic structure and genre or style. For example, Sample 1
starts with a phrase in Mozart’s style, then continues with a phrase in Walton’s style perhaps, which then turns into Scott Joplin$\ldots$ Sample 2
uses the harmonic language of a late Mahler symphony, along with the rhythmic language of a free jazz improvisation (I couldn’t make a time signature out of this clip). Sample 3 
starts with a phrase that could be the opening of a Romantic composition, and then takes off with a rhythmic structure that resembles a Bach composition, while keeping the Romantic harmonic language. 
Sample 4
is the most consistent of all. It sounds like a composition in the style of one of the Romantic piano composers (such as Liszt perhaps) and remains in that style throughout the clip. \\
\end{quote}
\vspace{0.25in}
\noindent {\bf Music Professor}: \\
\begin{quote}
[$\ldots$]{\it  {I'd guess human} because of a couple of ``errors'' in there, but maybe the AI has learned to throw some in! [$\ldots$]}\\
\end{quote}
\vspace{0.25in}
\noindent {\bf Pianist, TV \& Film Composer}: \\
\begin{quote} %
 Sample 1: resembles music in the style of Robert Schumann's Kinderszenen or some early romantic salon music. I'm fond of the rest after the little initial chord and melody structure. The tempo slows down slightly before the rest which sounds really lively and realistic - almost a little rubato. Then the distinct hard attack. Nice sense of dynamics. Also nice ritardando at the end of the snippet. Not liking the somewhat messy run but this almost seems as if someone had to study a little bit harder to get it right - {\it{it seems wrong in a human way.}} 
\end{quote}
\begin{quote}  %
Sample 2: reminds me of some kind of Chopin waltz, rhythm is somewhat unclear. The seemingly wrong harmony at the beginning seems to be a misinterpretation of grace notes. The trill is astonishing and feels light and airy. 
\end{quote}
\begin{quote}  %
Sample 3: Could be some piece by Franz Schubert. Nice loosely feeling opening structure which shifts convincingly into fierce sequence with quite static velocity. This really reminds me of Schubert because Johann Sebastian Bach shines through the harmonic structure as it would have with Schubert. Interesting effort to change the dynamic focus from the right to the left hand and back again.
\end{quote}
\begin{quote}
This is really interesting!
\end{quote}
\vspace{0.25in}
\noindent {\bf Piano Teacher}: \\
\begin{quote}
Sample 1:  Sounded almost Bach-like for about the first bar, then turned somewhat rag-timey for the rest
\end{quote}
\begin{quote}
Sample 2: Here we have a {\it{very drunken Chopin}}, messing around a bit with psychedelics
\end{quote}
\begin{quote}
Does that help at all?  Also, what do you mean by a regular piano sample library?  {\it{Did you play these clips as composed by the AI system?}}
\end{quote}
\medskip
\vskip0.25in

Overall, we note that the comments were quite consistent in terms of perceiving a human quality to the performance. Indeed, even though we made an effort to explain that all aspects of the MIDI file were generated by the computer, some people still wanted to double check whether in fact these were human performances.

While acknowledging the human quality of the performances, many of the musicians also questioned the strength of the long-term compositional structure. Indeed, creating music with long-term structure (e.g. more than several seconds of structure) is still a very challenging problem.

Many musicians identified the `style' as the mix of classical composers of which the data indeed consisted.

\section{Conclusion}

We have considered various approaches to the question of generating music, and propose that it is currently effective to generate in the space of MIDI performances. We describe the characteristics of an effective data set for doing so, and demonstrate a system that achieves this quite effectively.

Our resulting system creates audio that sounds, to our ears, like a pianist who knows very well {\it how} to play, but has not yet figured out exactly {\it what} they want to play, nor is quite able remember what they just {\it played}. Professional composers and musicians have provided feedback that is consistent with the notion that the system generates music which, on one hand, does not yet demonstrate long-term structure, but where the local structure, e.g. phrasing, dynamics, is very strong. Indeed, even though we did not frame the question as a Turing test, a number of the musicians assumed that (or asked whether) the samples were performed by a human.

\section{Acknowledgments}

We gratefully acknowledge all of the musicians who provided feedback on the samples. 
We thank members and visitors at Google Brain and specifically the Magenta team for discussions, including Adam Roberts, Anna Huang, Colin Raffel, Curtis Hawthorne, David Ha, David So, Fred Bertch, George Dahl, Jesse Engel, Kory Mathewson, Kyle Kastner, Natasha Jaques and Tim Cooijmans. Finally, we thank the reviewers for their useful feedback.\\

\bibliographystyle{plain}
\bibliography{references}

\end{document}